\documentclass[AMA,STIX1COL]{WileyNJD-v2}

\articletype{Article Type}%

\received{}
\revised{}
\accepted{}

\begin{document}

\title{Misconceptions, knowledge, and attitudes of secondary school students towards the phenomenon of radioactivity}

\author[1]{A.I. Morales*}

\author[2]{P. Tuz\'on}

\authormark{A.I. Morales \textsc{et al}}

\address[1]{\orgdiv{Instituto de F\'isica Corpuscular}, \orgname{CSIC-University of Valencia}, \orgaddress{\state{Valencia}, \country{Spain}}}

\address[2]{\orgdiv{Departamento de Did\'actica de las Ciencias Experimentales y Sociales}, \orgname{University of Valencia}, \orgaddress{\state{Valencia}, \country{Spain}}}

\corres{*Anabel Morales L\'opez, Instituto de F\'isica Corpuscular, C/Catedr\'atico Jos\'e Beltr\'an, 2, E-46980 Paterna, Valencia (Spain) \email{Ana.Morales@ific.uv.es}}


\abstract[Summary]{Since its serendipitous discovery in 1896 by Henry Becquerel, radioactivity has called the attention of both the scientific community and the broad audience due to its intriguing nature, its multiple applications and its controversial uses. For this reason, the teaching of the phenomenon is considered a key ingredient in the path towards developing critical-thinking skills in many secondary science education curricula. Despite being one of the basic concepts in general physics courses, the scientific teaching literature of the last 40 years reports a great deal of misconceptions and conceptual errors related to radioactivity that seemingly appear regardless of the context. This study explores, for the first time, the knowledge status on the topic on a sample of N=191 secondary school students and Y=29 Physics-and-Chemistry trainee teachers in the Spanish region of Valencia. To this aim, a revised version of a diagnostic tool developed by Martins \cite{Mar92} has been employed. In general, the results reveal an evolution from a widespread dissenting notion on the phenomenon, which is staunchly related to danger, hazard and destruction in the lowest educational levels, towards a more rational, relative and multidimensional perspective in the highest ones. Furthermore, the great overlap of the ideas, emotions and attitudes of the inquired individuals with the main misconceptions and conceptual mistakes reported in the literature for different educational contexts unveils the urgent need to develop new teaching strategies leading to a meaningful learning of the associated nuclear science concepts.}

\keywords{radioactivity, nuclear science, diagnostic study, misconceptions, conceptual errors}


\maketitle


\section{Introduction}
\label{sec:intro} 

The \emph{public awareness of Science} is an ambitious goal to be achieved by contemporary society. It implies informed, responsible, and active citizens in connection with the scientific and technological progress currently driving social change. Within this point of view, \emph{scientific literacy} \cite{Hol09,Ric09} is one of the key competences that students must develop during their compulsory education. According to some authors \cite{She75}, there are three levels of scientific literacy, namely practical, civic and cultural. While practical literacy refers to the ability of an individual to cope with daily scientific and technical issues such as food or health, civic literacy goes one step further. It reflects a higher level of involvement with science and technology. Among other aspects, citizens with a civic level of literacy have the ability to discern the scientific information provided by the social media, to critically judge the different scientific viewpoints, to understand the social, ethical and environmental impact of the scientific and technical activity and participate consequently in the related collective decision-making processes, to understand the nature of scientific research, and to appreciate the extent to which science and technology currently affect social welfare \cite{Mil83}. Meanwhile, cultural literacy requires the most sophisticated understanding of science and technology, as individuals can appreciate the scientific activity as a major intellectual one, responsible for the achievement of some of the most prominent milestones in the history of humanity. One essential aim of physics education research is reaching the second level of scientific literacy, the so-called civic literacy, in secondary education. Such a challenging task does not only imply the development of intellectual capabilities but also attitudinal, societal, and interdisciplinary skills \cite{Hol09}. In this respect, teaching science and technology in their historical, cultural, ethical, economical and political contexts \textendash i.e., teaching the relationships between Science, Technology, Society and Environment (STSE)\textendash~ has many benefits for the teaching-learning process, as discussed elsewhere \cite{Sol97,Ped03,Sol03}. 

Nuclear Science (NS) represents one of the most significant paradigms of STSE education \cite{Tsa13,Lev17}. The term `nuclear' is often related to contamination, catastrophes and weapons. It is worth remembering, though, that all matter around us is made up of atoms, and that inside every atom there is a nucleus that represents more than 99.9\% of the atomic mass. This feature makes nuclear properties crucial in the quest for the very essence of matter \cite{Har15,Kor18,But19}. One of them, radioactivity \textendash the spontaneous decay or disintegration of an unstable atomic nucleus accompanied by the emission of radiation\textendash~, is at the foundation of many useful or beneficial every-day applications in medicine, energy, industry, etc., yet it also entails potential risks for environment and humankind in the form of nuclear weapons, accidents, etc. When asked, the broad audience tends to relate NS with a harmful use of radioactivity. It is hence worth exploring the reasons behind the unpopularity of this term before designing any educational action on NS. As a matter of fact,  the social environment and the mass media are well known to stimulate the development of misconceptions and conceptual errors \textendash this is, ideas that do not fit in with currently accepted scientific theories\textendash~ in the population. Based on this, radioactivity-related misconception are very likely to differ at varying socio-demographic conditions, but surprisingly, the available science teaching literature shows resembling thoughts irrespective of the time and place of the reported studies. In particular, the only systematic survey carried out in the Spanish educational context dates back to the 80's  \cite{Pos89}. In their study, Posada and Prieto  \cite{Pos89} inquired 334 secondary education students in Madrid (Spain), reaching the conclusion that their perceptions on radioactivity were far away from the scientifically accepted theories.  
Apart from the long time elapsed from the study of Posada and Prieto, it is worth noting that the last concretion level of the educational curricula in Spain is defined by the individual autonomous regions. Hence, the implementation of new studies, better adapted to presently existing educational contexts in the country, becomes essential. In this manuscript, we approach the matter at hand in the educational reference frame of the Valencian Community, where the concept of radioactivity is approached for the only time in compulsory secondary education in the last course. To this aim, we use a diagnostic study based on a  work published elsewhere \cite{Mar92}. A universe of analysis of 220 secondary education students and trainee teachers has been used to discern the mental paths leading to the understanding of the phenomenon of radioactivity and its applications. Based on the answers obtained, we investigate how the mental representations at work influence the development (or mitigation) of misconceptions and conceptual errors for different educational levels. It is hence possible to track how education modifies and remodels these thoughts. 

\section{Literature review}
\label{sec:theo} 


Following the above argument, we provide a brief review of what educational research has hitherto found on the teaching-learning process of radioactivity. The starting point are the misconceptions and the conceptual mistakes about NS concepts reported in the literature. It is to note that the microscopic nature of radioactivity makes it an imperceptible phenomenon to human senses and, as a consequence, all erroneous ideas about radioactivity have an instructional origin, be it academic or not \cite{Ske93}. Moreover, its abstract and stochastic nature complicates more the task of reaching a meaningful learning of it. To achieve this goal, some authors suggest the use of analogies with more familiar processes as teaching resources \cite{Sch97,Lav17}. Even though, the limitations of the analogy should be properly identified beforehand, and the similarities and differences between the new concept and the known one clearly specified in order to avoid the appearance of erroneous ideas. 

The construction of misconceptions and conceptual mistakes has a complex an varied origin. Apart from the inadequate establishment of analogies, there are other factors fostering their appearance, such as the use of everyday words with a loose meaning, the construction of erroneous reasonings, etc. Furthermore, our anthropocentric view of the world and our limited sensory capabilities help biasing our perception of some processes and phenomena \cite{Dri85,Har00,Har03}. In the teaching-learning process of NS, the mass media and the teacher's interventions provide two of the most common means of misconceptions' transmission \cite{Nak06}. According to the constructivism theory \cite{Hen03}, such erroneous ideas will more likely be superseded by the correct ones if they are adequately taken into account in the instructional programs \cite{Nes87}. To this aim,  some authors have carefully analysed the misconceptions and learning difficulties of secondary-school students in different contexts and countries \cite{Pos89,Eij90,Mil96,Nak06,Neu12,Tsa13}. Some of the most noteworthy studies are contextualized in the framework of nuclear catastrophes \cite{Eij88,Mar92,Neu13,Plo19}, while others are limited to the cognitive confusion caused by the NS concepts themselves \cite{Mil96,Tsa13}. In Austria, Neumann and Hopf \cite{Neu12} studied the ideas about radiation held by students on their last compulsory year of secondary education. To this aim, they carried out a series of semi-structured interviews that brought students face to face with the alternative ideas most commonly reported in the scientific literature. They noticed that their perceptions were far away from those scientifically accepted, concluding that such ideas must be adequately approached by teachers in the Physics and Chemistry lessons. On the other hand, Tsarpalis et al. \cite{Tsa13} performed a comparative study with first-year university students in Greece and Turkey to inspect the misconceptions and emotions held after taking the compulsory secondary education training on radioactivity. This research was based on two newspaper articles that reflected the economical, ethical, and socio-political impact of nuclear applications on a global scale. As a result, they concluded that the Greek and Turkish secondary education programs were insufficient at the time to achieve a meaningful learning of the subject. Similar results were obtained by Nakiboglu and Tekin \cite{Nak06} with a  sample of Turkish students after receiving their Nuclear Chemistry course. Meanwhile, Colclough et al. \cite{Col11} analysed the knowledge and attitudes of trainee teachers about nuclear radiation associated risks, concluding that their competences need to be improved. Other authors, as Powell et al. \cite{Pow94} and Williams \cite{Wil95}, highlighted the need of introducing the historical, social and political frame in the programmed NS training in order to properly integrate STSE topics of common interest such as nuclear energy or nuclear waste disposal. On the other hand, Alsop \cite{Als01} stressed the relevance of the social and affective dimensions beyond the intellectual one in radioactivity-related questions. To this aim, the author carried out a comparative study between recently graduated students living in areas affected by high levels of radon gas and areas with normal levels of the radioactive gas. The first group seemed more informed and more objective with the effects of an overexposure to ionizing radiations than the second.      

\subsection{Reported misconceptions and mistakes}
\label{sec:theo} 

In summary, the educational research literature on the phenomenon of radioactivity and its related processes reports a series of misconceptions and conceptual mistakes that students of different countries have manifested over the past 40 years. Among the most common misconceptions, we find:

\begin{enumerate}
\setlength\itemsep{0.0em}

\item Radiation can accumulate in matter. According to Ref. \cite{Eij90}, after the Chernobyl accident a broad audience believed that radiation had entered into the food chain through the vegetal matter, which would have been expelled again after being absorbed and accumulated there.
\item Radioactivity is harmful for living beings \cite{Mil96}. As a result, there is a widespread fear to any type of radiation and in any situation. 
\item Radiation is highly destructive and dangerous. As indicated elsewhere \cite{Lin90,Ses10,Est12}, the main actors for this misconception to spread out are the Internet and other mass media.
 \item Radioactivity has a different effect on living and inert matter. Ref. \cite{Kla90} reveals that many students think living matter is more vulnerable to radioactivity than inert matter. Some of them use verbs such as `attract' and `absorb' to justify this thought, which is related to the analogy detected by Eijkelhof et al. \cite{Eij90} between radioactivity and a viral or microbial disease. 
 \item Objects and living beings exposed to radiation become radioactive. It is to note, though, that such a misconception may be certain when the radiation carries enough energy to excite the atomic nucleus or induce a nuclear reaction \cite{Plo16}.
 \item Radioactivity is conserved. This idea of conservation refers to the fact that many students think that if a body becomes radioactive, it remains radioactive forever \cite{Mil90}.
 \item Radioactivity is artificial. In fact, only a reduced number of students are aware of the existence of natural sources of radioactivity \cite{Neu12}. On the contrary, a vast majority thinks radioactivity can only be produced artificially in nuclear power plants \cite{Boy94}.
 \item Atoms cannot change their nature \cite{Nak06}. This thought is clearly in opposition to the scientific fact of spontaneous $\beta$ and $\alpha$ decay.
 \item Ionizing radiations are the cause of some CO$_2$ related environmental problems, such as greenhouse effect, pollution or the hole of the ozone layer \cite{Boy94,Neu12}.
\item All electrical devices emit harmful radiation. In their study, Neumann and Hopf \cite{Neu12} noted that some students were usually asked by their parents to  shut down the mobile phone and other electrical devices before going to sleep.  
\item Radiation can be used to detect feelings \cite{Neu12}. This is, perhaps, the most surprising of all misconceptions. The authors report that some students relate radioactivity with an esoteric aura.  
\end{enumerate}        

Conceptual mistakes have also been thoroughly detected in the scientific literature. The most common ones are:

\begin{enumerate}
\setlength\itemsep{0.0em}
\item The terms `radioactivity', `radiation' and `radioactive material' are often mixed up and ambiguously used. Kaczmarek, Bednarek and Wong \cite{Kac87} reported an amazing belief of second-year medicine students with no previous radiological training in New York University (US). Almost 75\% of them thought that objects in an X-ray room could become radioactive after a diagnose examination. Similarly, Prather \cite{Pra05} noticed that students attributed the same properties to ionizing radiation and radioactive material. Other studies \cite{Ses10} point to the interchangeable use of these terms in the mass media as the most likely error source.  
\item The terms `irradiation' and `contamination' are indistinguishably used. This is most likely a consequence of the interchanged use of the words `radiation' and `radioactive material', as the first is related to irradiation and the second to contamination as stated in Ref. \cite{Mil96}.
\item The terms `isotope', `radioisotope', `atom', and `chemical element' are often confused or vaguely differentiated \cite{Nuc18}. 
\item Nuclear fusion and fission reactions are usually confused. Indeed, many students think of fission as the only existing nuclear reaction \cite{Tsa13}. 
\end{enumerate}

More conceptual mistakes related to NS concepts such as the atomic mass, the atomic number, the half-live, etc. can be found in the literature \cite{Nak06}. Nonetheless, these are identified \emph{after} teaching interventions at undergraduate level and thus fall out of the scope of this work. Instead, we will focus on the ontological and phenomenological understanding of the phenomenon of radioactivity, as well as on the emotions, attitudes and interests that radioactivity stimulates in secondary education students.

\section{Methodology}
\label{sec:meth}

The starting point of this work is a questionnaire designed and validated in the doctoral work of Martins \cite{Mar92} to explore how Brazilian secondary school students assimilated the information received about radioactivity in a context related to the accident of Goi\^ania. The reference work approaches the phenomenon of radioactivity from three common dimensions in science education research, namely the theoretical (\emph{What is radioactivity?}), the educational (\emph{How radioactivity can be taught?}) and the social (\emph{Which emotions inspires radioactivity?}) using two main working tools, a questionnaire and a personal interview. The part relating to the questionnaire is mainly focused on the sources and nature of knowledge about radioactivity, i.e., how do students understand the entity itself, its properties and where this knowledge has been acquired. 

In the present work, the diagnostic tool used is an updated questionnaire adapted from that provided in Ref. \cite{Mar92}. The new questionnaire has been complemented with open-ended questions to obtain detailed information on the misconceptions and conceptual errors held by a sample of secondary school students and trainee teachers in the Spanish region of Valencia. Moreover, some of the original close-ended questions have been shortened in order to finish the survey in approximately one class session (40-50 min). The universe of analysis are 191 secondary school students and 29 Physics-and-Chemistry trainee teachers. In both cases, information on socio-demographic variables has been collected. In the secondary school sample, these are `level', `group', `age' and `sex', while in the master sample they are `career', `age' and `sex'. In total, a matrix with 220 cases has been generated. Such a comprehensive sample allows one to evaluate the influence of the academic training not only in the mental construction of the concept of radioactivity, but also in the development of alternative ideas, conceptual errors, attitudes and interests.

\subsection{Field work}
\label{sec:field}

The study has been carried out in a secondary school and a public university, both located in the metropolitan area of the city of Valencia (Spain). The secondary school sample consists of 10 groups of students aged between 13 and 19 taking the 2$^{nd}$, 3$^{rd}$ and 4$^{th}$ courses of compulsory secondary education (ESO, in Spanish acronym) and the first optional course of secondary school (called 1$^{st}$ Bachillerato). According to the Physics-and-Chemistry Spanish and Valencian curricula, NS concepts such as \emph{atomic nucleus} and \emph{isotope} are introduced for the first time in 2$^{nd}$ ESO, while the phenomenon of radioactivity is approached for the first and only time during the compulsory education in the $3^{rd}$ course of ESO. In this level, the concepts of  $\alpha$, $\beta$, and $\gamma$ radiation are introduced. The social and civic dimensions of NS are also  integrated through the analysis of some applications of radioactivity, such as the nuclear energy and the controversial problem of the radioactive waste disposal. NS is resumed in the second optional course of secondary school (called 2$^{nd}$ Bachillerato). This means that only students opting for the scientific and technical itinerary in their last optional course of secondary education will treat in detail NS aspects basic to achieve the civic level of scientific literacy. Concomitantly, concepts with a certain connection to NS; such as the atom, the atomic structure, its composition and the atomic interactions, are widely approached in all secondary levels.  

The university sample is comprised by 29 trainee teachers attending a master on secondary school teaching on Physics and Chemistry, aged between 22 and 35. Most of them have a degree in Chemistry, although other career backgrounds such as Biochemistry, Biotechnology, Biology, Engineering and Physics can be found. Around 40\% of them have received advanced training on NS at undergraduate level, and nearly 30\% has or is about to have a Ph.D in any of these science disciplines.

\begin{table*}
  \centering
  \caption{Rubric showing the level of adequacy, scores and correction criteria for the open-ended questions listed in Item 6 of the supporting information.}
   \begin{tabular}{p{10em} | l | c | p{25em}}
    \hline
    \hline
     \textbf{Question}  & \textbf{Level}  & \textbf{Score} & \textbf{Correction criteria} \\ 
\hline 
   \multirow{3}{10em}{Q6.1) How many types of nuclear radiation do you know?} & Good  & 0,5 & $\alpha$ rays, $\beta$ rays and $\gamma$ rays     \\
    & Satisfactory   & 0,25 & Naming only one or two of the three most common types of nuclear radiation  \\
    & Poor  & 0   &  Failing to cite any type of nuclear radiation  \\
    \hline
    
      \multirow{3}{10em}{Q6.2) Which one do you think is the most dangerous for human beings?} & Good  & 0,5 & $\gamma$ rays as they are more penetrating / Depends on the distance to the source, the source intensity and the energy and type of nuclear radiation \\
   & Satisfactory   & 0,25 & Failing to justify which type of nuclear radiation is the most dangerous \\
   & Poor   & 0 & Providing an incorrect answer \\
      \hline
    
     \multirow{3}{10em}{Q6.3) Why some nuclei are radioactive?} & Good  & 0,5 & Because they are (energetically) unstable / Because the combination of protons and neutrons is not stable \\
   & Satisfactory   & 0,25 & Mentioning the `composition' of the radioactive substance without explicitly speaking about neutrons and protons\\
   & Poor   & 0 & Providing an incorrect answer \\
      \hline
      
          \multirow{3}{10em}{Q6.4) How can we protect ourselves from a radioactive substance?} & Good  & 0,5 &  Moving away from it / Letting it decay through time / Putting adequate material before to stop the emitted radiation (e.g., radiation suits) \\
   & Poor   & 0 & Providing an incorrect answer \\
      \hline
 
          \multirow{3}{10em}{Q6.5) How many applications of radioactivity do you know? List them.} & Good  & 0,5 &  Citing at least two correct applications. \\
   & Satisfactory   & 0,25 & Citing only one correct application / Citing correct and incorrect applications \\
   & Poor   & 0 & Providing an incorrect answer \\
      \hline
      
          \multirow{3}{10em}{Q6.6) Do you think radioactivity affects living and inert matter equally?} & Good  & 0,5 &  No, because it has health effects for living matter / Yes, because nuclear radiation can induce nuclear reactions in all matter, be it living or inert\\
   & Poor   & 0 & Answer `yes' or `no' without providing any justification / Providing an incorrect answer \\
      \hline
      
            \multirow{3}{10em}{Q6.7) Can radioactivity turn objects radioactive?} & Good  & 0,5 &  Yes, if it induces a nuclear reaction with a final radioactive product / Yes, objects become radioactive in areas affected by nuclear catastrophes / Yes, if the radioactive material is ingested or adhered / Yes, by exposure to nuclear radiation \\
   & Poor   & 0 & Answer `yes' or `no' without providing any justification / Providing an incorrect answer \\
 
       \hline
       \hline
    \end{tabular}%
  \label{table1}%
\end{table*}%


 \subsection{The questionnaire}
 
The questionnaire used here for data collection is provided as supporting information. Items 1-5 have been extracted from Ref. \cite{Mar92}, while Item 6 is new and consists of a series of open-ended questions that aim at better knowing the individual ideas of the inquired students about radioactivity. Item 1 explores the ontological nature of the phenomenon. To this aim, different entities, related to everyday objects, physical concepts and processes are listed in a table together with two possible answers: `yes', if the surveyed student believes the entity is somewhat related to radioactivity, and `no' in the opposite case. It should be noted that the list is exactly equal to that provided in Ref. \cite{Mar92}, except for the entity \emph{particles}, which has been added in the new version. Item 2 inspects the properties attributed to radioactivity. A table listing pairs of adjectives with opposite meanings, each one marked as (a) or (b), is provided. Four answers allow inquired subjects to consider the extent to which radioactivity can be defined with the pair of adjectives: `totally or possibly (a)', `(a) and (b)', `nor (a) nor (b)', and `totally or possibly (b)'. The list is very similar to that designed by Martins \cite{Mar92}, except for several pairs of adjectives that have been eliminated, mainly because the misconceptions or ideas behind them are treated in other questions. The aim of Item 3 is discerning the own perception of individuals about their knowledge on radioactivity. The five answers provided in Ref. \cite{Mar92} are also given here: `null', `scarce', `superficial', `reasonable', or `deep'. Meanwhile, Item 4 examines the main sources of information of the inquired students. The  items `home' and `Internet' have been added to the list of answers provided by Martins \cite{Mar92} (`school', `TV', `newspapers/magazines', `other'). Item 5 provides a list of NS applications and issues (medicine, energy, food, etc.) to explore which stimulate more interests in individuals.  Finally, Item 6 consists of seven open-ended questions with a twofold aim. The first is fostering the emergence of misconceptions and conceptual mistakes, and the second is figuring out how well the NS contents learned in secondary education have been implemented. With this latter purpose, Item 6 has been evaluated numerically using two scoring criteria. In the first one, answers are rated as `poor' (0), `satisfactory' (0.25) and `good' (0.5); while in the second, generally used for dichotomous questions, the answers are only rated as `poor' (0) and  `good' (0.5). The correction criteria are provided in Table \ref{table1}.

\section{Results and discussion}

The collected data have been treated statistically with the free analysis program R \cite{R}. In the following, we will discuss the results separately for every item of the questionnaire.   

 \begin{figure}[t]
\centerline{\includegraphics[width=0.6\textwidth]{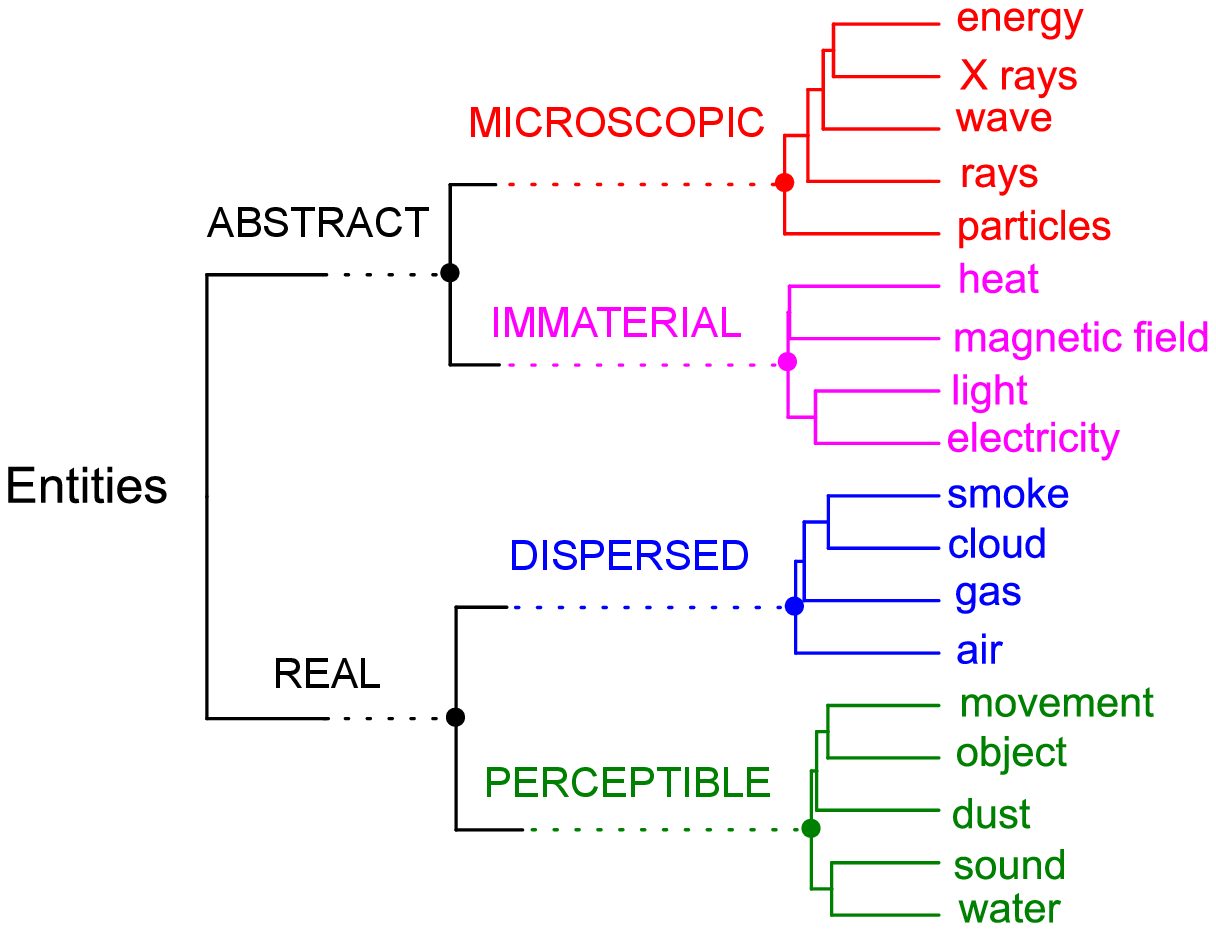}}
\caption{Dendrogram showing the hierarchical cluster analysis results of Item 1 for the secondary school sample. Details are discussed in the text.}
\label{fig1}
\end{figure}  

\begin{table}
  \centering
   \renewcommand{\tabcolsep}{0.30 cm}
   \caption{Results of the Fisher tests for the entities provided in Item 1. These are ordered in columns as a function of the ontological categories obtained in the cluster analysis.}
    \begin{tabular}{l l| l l | l l| l l}
    \hline
    \hline
    \multicolumn{2}{c|}{Perceptible} & \multicolumn{2}{c|}{Microscopic} & \multicolumn{2}{c|}{Dispersed} & \multicolumn{2}{c}{Immaterial} \\
    \hline
    \hline
    Entity & Fisher test & Entity & Fisher test  &   Entity & Fisher test & Entity & Fisher test \\
    \hline
    Water  & 0.2   & Particles & 0.084 &   Air  & 0.13  & Electricity & \textbf{0.0019}  \\
    Object & 0.33  & Rays & 0.58 &    Gas   & \textbf{0.0013} & Light   & 0.21 \\
    Dust & 0.46  & Wave  & 0.53 &   Cloud  & 0.36  & Magnetic field & 0.093 \\
    Sound & 0.28  & X rays & 0.29 &     Smoke  & \textbf{0.0017} & Heat & 0.28\\
    Movement & 0.21  & Energy & 0.034 &  & & & \\
    \hline
    \hline
    \end{tabular}%
  \label{table2}%
\end{table}%

\begin{figure*}[t]
\centerline{\includegraphics[width=\textwidth]{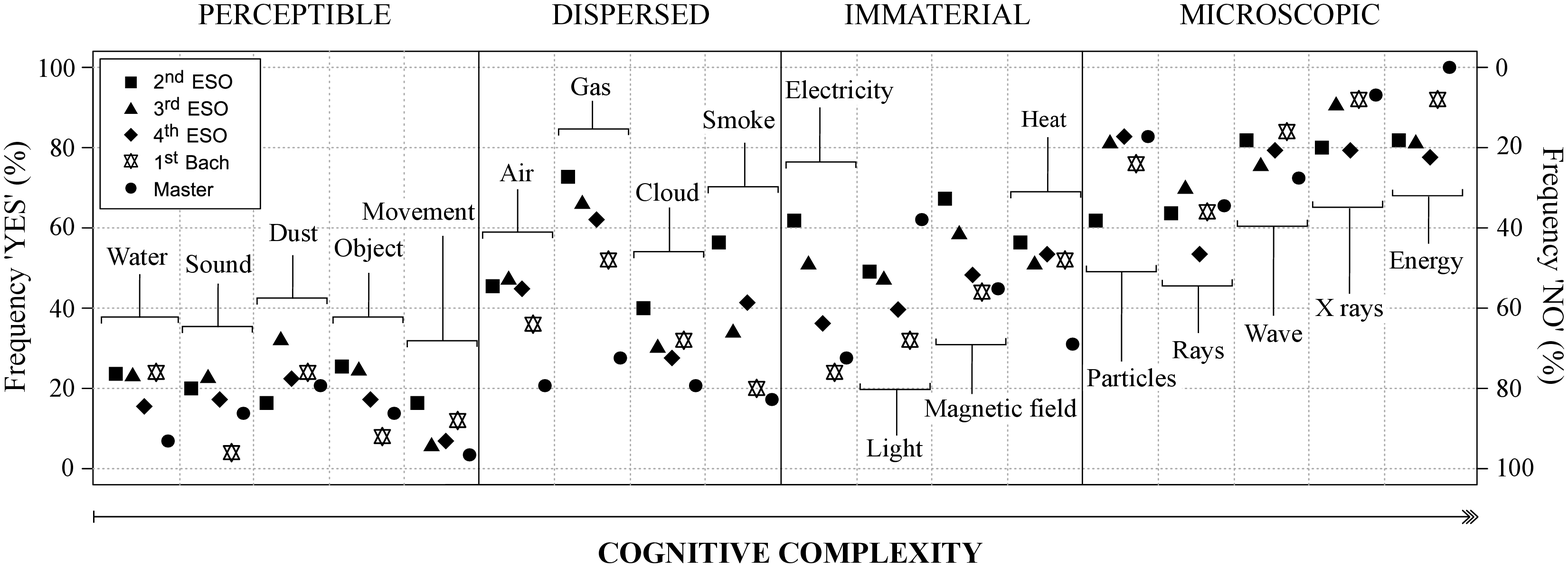}}
\caption{Response distributions per academic level (see legend) for Item 1. The frequencies of positive answers are indicated in the left axis and those of negative answers in the right axis. The entities are displayed as a function of the ontological categories identified in the cluster analysis: `perceptible', `dispersed', `immaterial', and `microscopic'.  Frequencies are normalized to the total number of students per academic level.}
\label{fig2}
\end{figure*}

\subsection{Item 1: On the nature of radioactivity}
\label{sec:nature} 

The first item explores the mental schemes built by students in relation to the idea of radioactivity. The aim is to classify how they connect this phenomenon to the physical concepts and processes shown in the first table of the supporting information.   
The answers to Item 1 have been analysed using the hierarchical clustering technique \cite{Rok05}. This procedure identifies separate groups of entities in order to provide a categorization of the considered elements. The dissimilitude between pairs of entities is evaluated using a metric that calculates the distance between them. Meanwhile, clusters are created from a linkage algorithm based on the selected metric. In our case, we have used the euclidean metric and Ward's clustering algorithm \cite{War63}. Results for the secondary school sample are shown in Fig. \ref{fig1}. Two main clusters or ontological categories can be clearly distinguished. The first comprises well-known concepts that are accessible to our senses and are present in our daily life, whereas the second includes intangible or immaterial entities, inaccessible to human senses and, hence, less familiar to individuals. For this reason, the first category has been called `real' and the second one `abstract'. In turn, the category `real' is subdivided in two additional clusters, `perceptible' and `dispersed'. Concepts such as water, object, dust, sound, and movement are comprised within the sub-category `perceptible' due to their accessibility to human senses. Meanwhile, air, gas, cloud and smoke are grouped in the sub-category `dispersed' for their dissipated and/or scattered nature, being all made up of invisible particles. The category `abstract' is also formed by two sub-categories, so-called `immaterial' and `microscopic'. Whilst the first refers to physical concepts related to thermodynamics and electromagnetism such as electricity, light, magnetic field, and heat, the second includes active and imperceptible entities related with atomic, nuclear, and particle physics such as particles, rays, waves, X rays, and energy. This sub-category defines the most complex cognitive group due to the high level of abstraction required for a deep comprehension of the concepts. 

It is interesting to examine the degree of correlation established by the inquired students between the concept of radioactivity and the four ontological categories identified in the hierarchical clustering analysis, as well as the influence of formal education in the construction of the concept's mental picture. With this aim, we have performed a categorical analysis of the answers to Item 1 for each of the academic levels described in Section \ref{sec:field}, including the trainee teachers.  An illustration is given in Fig. \ref{fig2}, where the entities are ordered in increasing cognitive complexity according to their category. The response distributions, normalized to the total number of students per level, are shown per academic course (see legend). The frequency of positive answers is indicated in the left axis, while that of negative answers is shown in the right one. 

At first sight, one can appreciate a gradual increasing trend of positive answers with the cognitive complexity of the categories. Whilst the percentage of positive answers is below 30\% for the first ontological group (`perceptible'), it exceeds the 60\% for the last one (`microscopic'). The observed evolution is in line with the progression from simple to complex, close to remote, familiar to strange, and definite to abstract that connects the discussed concepts. Hence, the surveyed individuals seem to use selection criteria such as the accessibility to human senses, the  immateriality, the complexity and the level of abstraction to make their choices. Within this perspective, radioactivity is assigned to the last ontological group, as it is appreciated as highly abstract, complex, remote and completely inaccessible to senses. In summary, it is perceived as an intricate entity very difficult to understand.

\begin{figure}
\centerline{\includegraphics[width=0.55\textwidth]{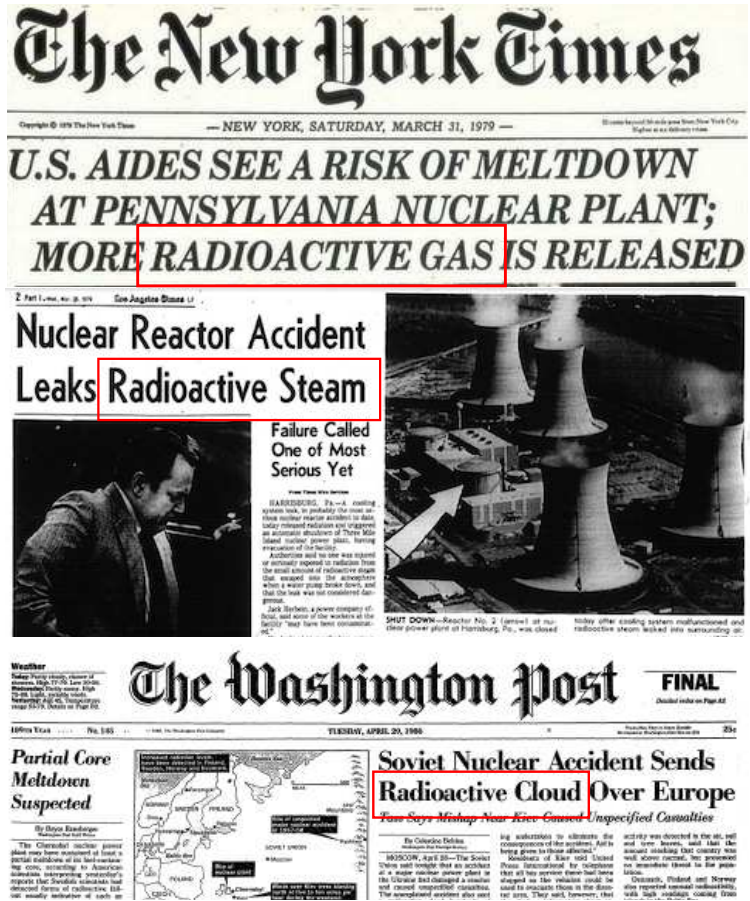}}
\caption{Newspaper's front pages with headlines about nuclear accidents that spread a `dispersed' conception of radioactivity.}
\label{fig3a}
\end{figure}

For each entity listed in Item 1, the dissimilarity of the response distribution among different educational levels has been evaluated with a Fisher test. The results are shown in Table \ref{table2}. The only entities that show statistically significant differences are `gas', `smoke' and `electricity'. These are concepts assigned to the intermediate categories `dispersed' and `immaterial'.  Interestingly, these cognitive groups, which are halfway through the most tangible and abstract categories, are the ones that generate more confusion in the compulsory secondary education. On the one hand, `dispersed' entities, such as gas or cloud, often appear linked to the term \emph{radioactive} in the mass media (see Fig. \ref{fig3a}). The use of such a loose vocabulary makes readers confuse the terms \emph{radiation}, \emph{radioactivity} and \emph{radioactive material} \cite{Ses10}. In the case of students attending compulsory secondary education, the lower level of formal training makes them more vulnerable to the assimilation of such misconceptions. As a result, they associate this cognitive group to radioactivity more often than students attending higher educational levels. On the other hand, the confusion caused by the group `immaterial' can be understood in terms of the abstract nature of radioactivity. The idea of this phenomenon as something intangible, invisible and complex is the most likely reason for a higher connection to physical entities such as electricity and magnetic field, as they are equally weightless, complex and vague to students with a low educational background.  In summary, formal instruction seems to play an important role in establishing the degree of similarity between the concept of radioactivity and the ontological groups `dispersed' and `immaterial'. On the contrary, the categories `perceptible' and `microscopic' raise a broadest agreement among the different educational levels. They represent the least and most linked with the idea of radioactivity, respectively. Unsurprisingly, the group `microscopic' is integrated by concepts of atomic, nuclear and particle physics.   

\subsection{Item 2: On the properties of radioactivity}
\label{sec:properties}

The second item explores which properties are more commonly attributed to radioactivity and how do these attributions evolve with the instructional level. Similarly to Item 1, a categorical analysis of the responses for each pair of adjectives has been performed. In order to ease the discussion, the pairs of adjectives have been grouped in five general qualities: microscopic nature, uniqueness, duration, activity, and functionality. The classification is shown in Table \ref{table3}. 
The corresponding response distributions are shown  in Figs. \ref{fig3} and \ref{fig4} for each of the academic levels discussed in Section \ref{sec:field}. As before, the frequencies are normalized to the total number of students per level. For the sake of clarity, the pairs of adjectives with opposite meaning are indicated  in the title of each graph with a label, `(a)' or `(b)'. The frequency of responses opting for adjective (a) are indicated by a yellow bar, while those opting for adjective (b) are shown in blue. The option `(a) and (b)' is indicated in green, while `nor (a) nor (b)' is shown in red. In the following, we discuss the distributions:

\begin{table}
  \centering
   \renewcommand{\arraystretch}{1.4}
   \renewcommand{\tabcolsep}{0.25 cm}
   \caption{Classification of the pairs of adjectives listed in Item 2 according to their assigned general quality.}
    \begin{tabularx}{\textwidth}{ c| cX}
    \hline
    \hline
    \bf{General quality} & \bf{Pairs of adjectives} \\
    \hline
          \multirow{2}{8em}{\it{Microscopic nature}}  & material-immaterial, amorphous-shaped, solid-liquid, perceptible-imperceptible, light-heavy, \\ 
          & divisible-indivisible, corpuscular-wavy, detectable-undetectable, measurable-immeasurable \\
    \hline
    \it{Uniqueness}  & common-special, natural-artificial, frequent-rare \\
    \hline
      \it{Duration} & transient-permanent, brief-lasting, stable-unstable, increasing-decreasing \\
    \hline
      \it{Activity}  & strong-weak, static-mobile, energetic-inert, controllable-uncontrollable \\
    \hline
        \it{Emotion and functionality} & destructive-creative, harmful-beneficial, dangerous-safe, useful-useless \\
           \hline
    \hline
    \end{tabularx}
    \label{table3}
\end{table}

\subsubsection{Microscopic nature}

This group of adjectives is shown in Fig. \ref{fig3}. The main aim here is to inspect the ideas about the structural nature of the particles involved in the process of radioactivity. In turn, these  allow one to better understand the causes behind the confusion caused by the terms `radiation' and `radioactive material', as only the latter can be described using macroscopic properties. In the case of the pair `material-immaterial', about 50\% of the surveyed students select the option `immaterial', in accordance with the results of the cluster analysis of Item 1 (see Section \ref{sec:nature}). However radiation can be both material and immaterial (green option) as it can consist of quantum particles with mass, such as $\alpha$ or $\beta$ particles, or without mass, such as photons. It is to note that the mental picture of immateriality persists in the higher educational levels; in particular, there is no increase of the green option [`(a) and (b)'] for the trainee teachers, even if a significant number of them have received advanced undergraduate training on NS. There is even a clearer consensus on the perception of radioactivity as an amorphous entity, most likely due to its microscopic and abstract nature. It is worth noting, though, that nuclei do have shapes \cite{Luc01}. In particular, radioactive nuclei typically show non-spherical shapes as a result of their neutron-to-proton asymmetry and, in some cases, even the coexistence of different shapes in the same nucleus has been reported \cite{Mor17}. The nuclear shape can be determined using nuclear techniques such as $\gamma$-ray spectroscopy \cite{Lis13,Dra16}. 

The adjectives `solid' and `liquid' can only describe a macroscopic material entity. For this pair of adjectives, the number of inquired students that select any of the three first options  [`(a)', `(b)', or `(a) and (b)?] fades away at increasing educational level, while the last option [`nor (a) nor (b)'] concomitantly increases. Thus, formal instruction seems to play a mitigating role on the main conceptual mistake associated to the teaching of radioactivity. Meanwhile, around 40\% of students in compulsory secondary education believe that radioactivity is `perceptible' (yellow). More surprisingly, 40\% of the master students select either this option or the green one (`perceptible and imperceptible'), even if radioactivity is a microscopic phenomenon completely inaccessible to human senses. Although the inquired subjects do not give any particular reason for their choices, cause-effect relationships might be at the heart of this misconception as the effects of radioactivity in living organisms are well known. 

The confusion caused by the microscopic composition of matter can be clearly appreciated in the response distributions of the pairs `divisible-indivisible' and `corpuscular-wavy'. They show the most varied responses in this group, bringing to light the lack of familiarity with the aspects of Modern Physics \cite{Fis92,Gil93,Tuz16}. In both cases, the correct answer [`(a) and (b)'] holds a small percentage for the two pairs of adjectives in secondary education. A pattern that is somehow expected, as the Valencian curriculum for these educational levels does not foresee formal instruction on quantum physics aspects such as the wave-particle duality. On the other hand, the response distributions of the master students are quite surprising. For the pair `divisible-indivisible', a scarce 20\% opts for the green answer. This percentage raises to 60\% in the pair `corpuscular-wavy', yet nearly 40\% of the trainee teachers select the answers `corpuscular' (yellow) or `wavy' (blue), evincing a retrieval of the classical picture of Physics to construct the mental concept of radioactivity. This suggests that the macroscopic representation of the microscopic world persists through formal education up to the highest instructional levels.

The idea of materiality is revisited again in the pair `light-heavy'. The preferred option of secondary school students is `light' (yellow), possibly making reference to radiation; while the frequency of the adjective `heavy', relating to radioactive material, is kept below 20\% for all levels. On the other hand, the percentage of subjects selecting `nor light nor heavy', in association with immateriality, is rather low except for the master course. This result is at variance with those for the pairs `material-immaterial' and `solid-liquid', and unveils a certain inconsistency in the mental schemes about radioactivity built by the students. Finally, the responses to the pairs of adjectives `detectable-undetectable' and `measurable-immeasurable' show an increasing frequency of option `(a)' with the educational level; this is, most of students correctly state that radioactivity can be detected and measured. Even so, the other options maintain a certain weight in the lower educational levels, most probably due to the connection established between the terms `perceptible', `detectable', and `measurable'.

\subsubsection{Uniqueness}

\begin{figure*}
\vspace{0.3cm}
\centerline{\includegraphics[width=0.9\textwidth]{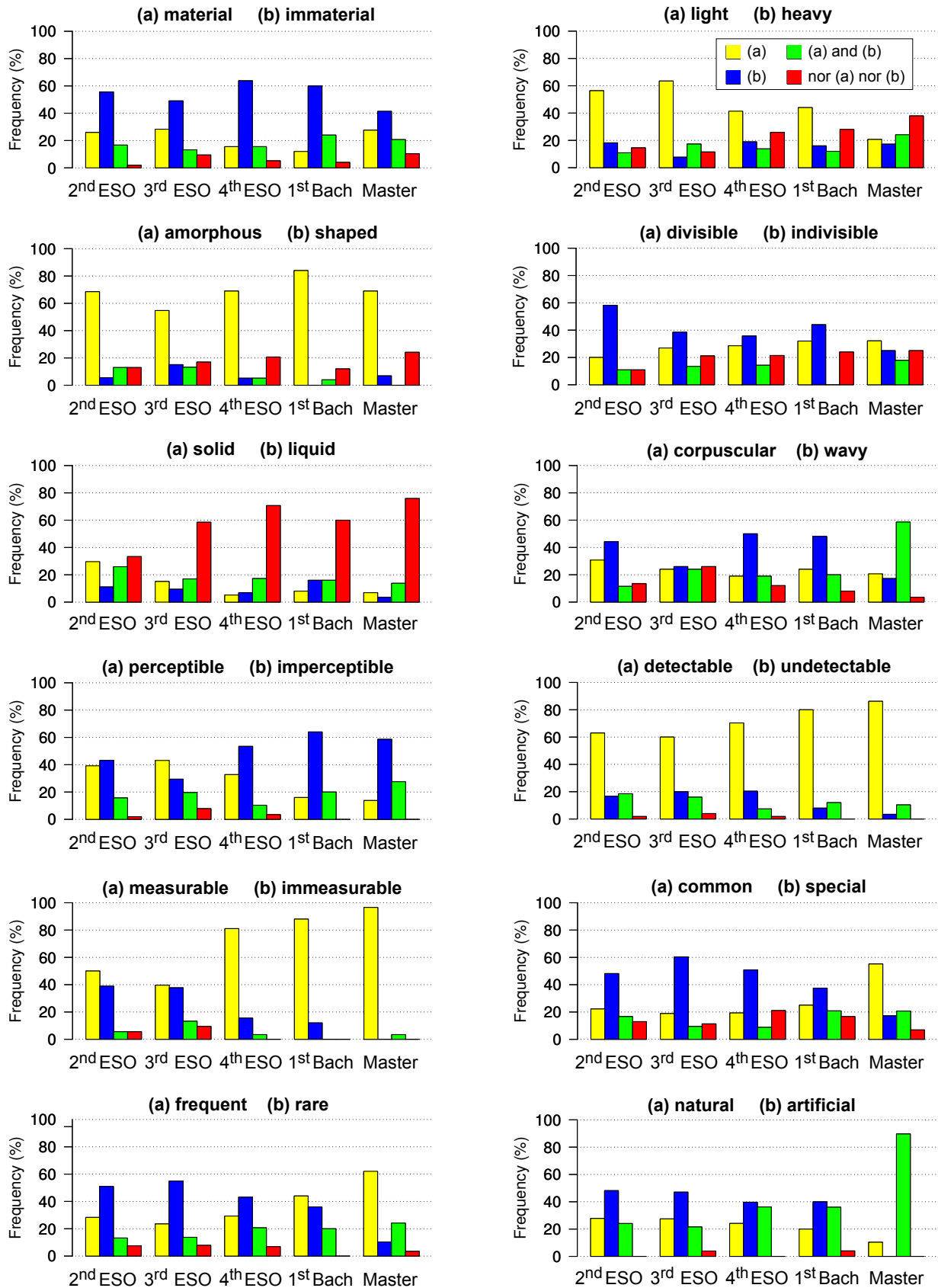}}
\caption{Answer distributions for the pairs of adjectives listed in Item 2. Frequencies are given for each educational level in increasing order. Labels (a) and (b) in the legend refer to each of the two adjectives labelled as (a) and (b) in the title of each graph.}
\label{fig3}
\end{figure*}

Another important issue about the mental picture of radioactivity is its uniqueness. This attribution is explored in Item 2 with three pairs of adjectives, `common-special', `natural-artificial', and `frequent-rare'. In the three cases, there is a clear difference between the compulsory secondary school students and the trainee teachers, being the most notorious one the pair 'natural-artificial' (see Fig. \ref{fig3}). Whilst the former mainly consider radioactivity as an `special', `artificial' and `rare' entity (blue), the later prefer to define it as `common' (yellow), `natural and artificial' (green) and `frequent' (yellow). Thus, the perception in compulsory secondary education seems closely linked to the misconception that radioactivity has an artificial-only origin \cite{Boy94}. Meanwhile the master group, with a higher level of instructional training, is perfectly aware of the existence of natural sources such as the cosmic rays or the Earth. The gradual increase of the correct options with the educational level is a clear indication that formal instruction does amend this misconception. According to these data, the conceptual bridge towards the correct natural picture of radioactivity seems to be built in the optional secondary levels. 

\begin{figure*}
\vspace{0.3cm}
\centerline{\includegraphics[width=0.9\textwidth]{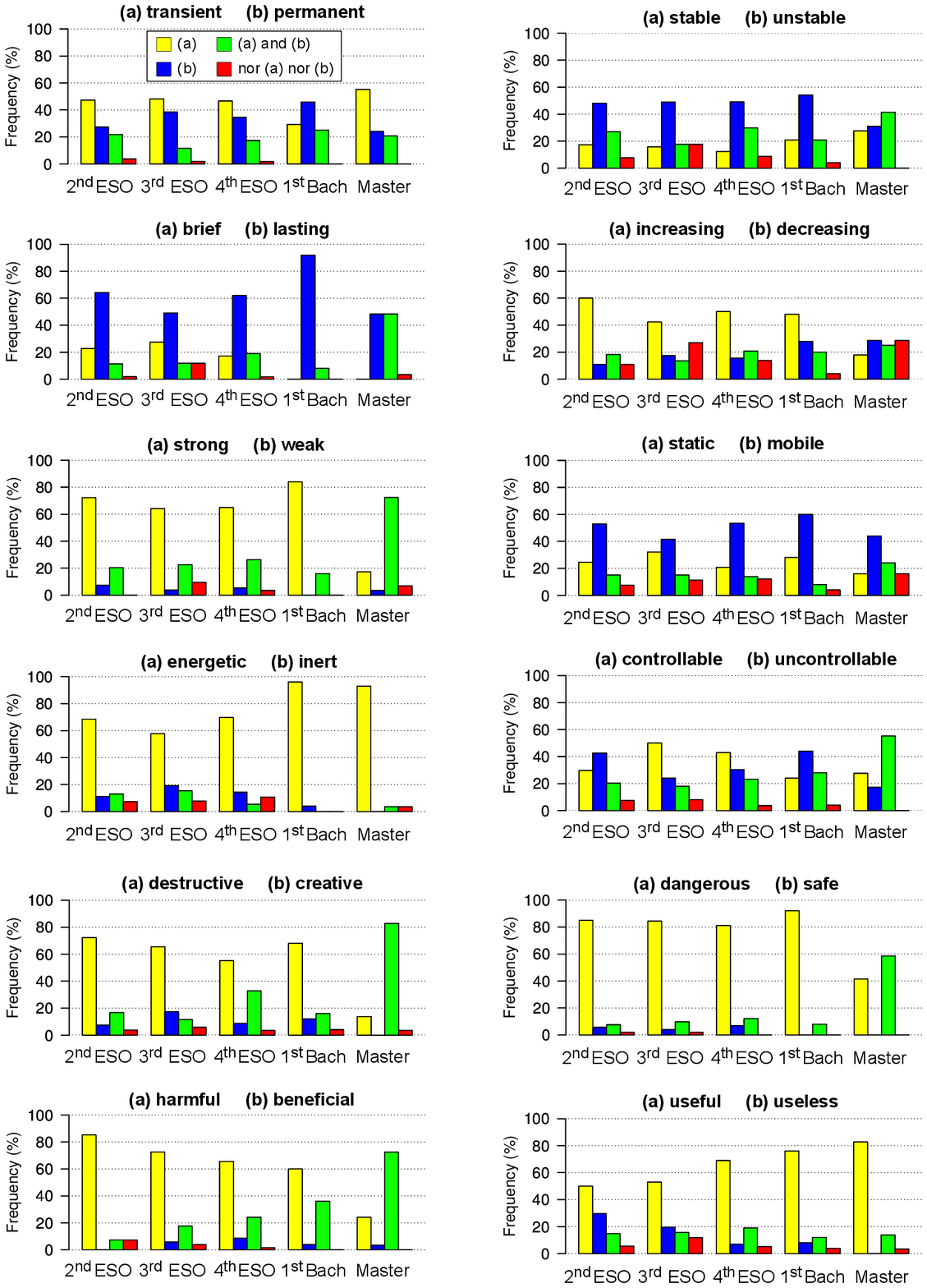}}
\caption{Answer distributions for the pairs of adjectives listed in Item 2. Frequencies are given for each educational level in increasing order. Labels (a) and (b) in the legend refer to each of the two adjectives labelled as (a) and (b) in the title of each graph.}
\label{fig4}
\end{figure*}  

\subsubsection{Duration}

This group of adjectives is shown in Fig. \ref{fig4}. The idea of radioactivity as a constant, perpetual entity is a well-known misconception in science education research \cite{Mil90}. The knowledge of the nuclear waste disposal problem or the long time needed to decontaminate areas affected by nuclear accidents might be at the heart of this misconception. It is to note, though, that the duration of radioactivity depends on the half-life, a property that is unique to each radionuclide and that can range from a few microseconds to thousands of years \cite{Nuc18}. Although it is true that radioactivity can temporarily increase for radioactive chains including more than one unstable nucleus, the radioactive decay law clearly establishes that the original sample of radionuclides decreases exponentially with time. In the present survey, secondary school students prefer to describe radioactivity as a `lasting' and `increasing' entity. At the same time, they opt for the adjectives `transient' and `unstable'. This clash of ideas (that might be partially contradictory) is manifested through the generally highly fragmented response frequencies. For instance, in the pairs `transient-permanent' and `stable-unstable', the answers `permanent' (blue) and `stable' (yellow) combined with the option `(a) and (b)' (green) amount to 40\% or more in all levels. Surprisingly, beyond 60\% of the trainee teachers opt for the answers `stable' or `stable and unstable', a  percentage that raises to nearly 100\% for the options `lasting' and `brief and lasting'. Even if they represent the group with the highest level of scientific literacy \textendash and hence should be the more coherent of all\textendash, this results seems in contradiction with those obtained for the pair `transient-permanent', for which nearly 60\% of them selected the option `transient'. It would be interesting to inspect the underlying mental schemes that lead them to connect radioactivity with the idea of stability.

\subsubsection{Activity}

We examine again the confusion generated by the terms `radiation' and `radioactive material', since the first is active and the second passive. In the case of the pair `strong-weak', a vast majority of inquired students in secondary school believe that radioactivity is a `strong' process (yellow, see Fig. \ref{fig4}). This perception changes abruptly for the master students, who consider it can be `strong and weak' (green). The consensus to define radioactivity as an `energetic' entity is even larger, reaching nearly 100\% in the higher levels. This answer is in complete agreement with \textendash and indeed confirms\textendash~ the results obtained for the cluster analysis of Item 1 (Section \ref{sec:nature}). There, the entity `energy' was located in the ontological group `microscopic', considered by students as the one conceptually closer to radioactivity.

Meanwhile, the pairs `static-mobile' and `controllable-uncontrollable' generate more confusion in students, leading to a higher fragmentation of the response distributions. In the first case, well beyond 40\% select the blue option (`mobile'), indicating that the main trend is establishing mental schemes in relation to radiation rather than to radioactive material. It is to note that the responses to this pair preserve a certain coherence with the pair 'light-heavy'. For instance, the preferred options of secondary school students are `light' and `mobile', clearly making reference to radiation. Meanwhile, the percentages of students selecting the options `heavy' and `static', both making reference to radioactive material, are quite similar. For the pair `controllable-uncontrollable', we see again a variety of views that only appears to partially clear up for the master group. In this case, nearly 60\% thinks radioactivity can be both, `controllable and uncontrollable' (green bar).

\subsubsection{Emotion and functionality}

The later group of adjectives is shown in Fig. \ref{fig4}. The aim here is to inspect the affective and emotional biases inspired by the idea of radioactivity. As expected, there is a widespread belief among secondary school students that radioactivity is `destructive', `harmful' and `dangerous' (yellow). Hence, the association of radioactivity with negative emotions and feelings reappears in this study. On the other hand, the options `destructive and creative' and `harmful and beneficial' (green) rise gradually up to about 80\% for the master group. This changing trend shows the effect of formal education on the emotional misconceptions about radioactivity. Definitely, the trainee teachers appear to be more aware of the complexity of radioactivity, which can be destructive or creative, harmful or beneficial, or dangerous or safe depending on several factors, such as the absorbed dose or the time of exposure. In the later case, though, the percentage of master students selecting the yellow option (around 40\%) is larger than for the two former pairs of adjectives (around 20\%). This suggest that the connection to danger persists more robustly through formal training than the associations to destruction and damage. Surprisingly, a growing majority in all educational levels still consider radioactivity `useful' (yellow). This result seems in disagreement with the \emph{emotional} pairs of adjectives. The conflict between the information given by academic and non-academic means seems to appear here. In order to mitigate these incoherences, it is worth to properly integrate the STSE relationships of NS in instructional programs \cite{Als01,Tsa13}.   

\vspace{-1cm}
\subsubsection*{}

From the categorical analysis of Items 1 and 2, one can conclude that radioactivity is perceived as an extremely complex phenomenon, far off experience, unfamiliar and highly abstract, comprised within the most advanced hierarchical category in the cluster analysis (see Fig. \ref{fig1}). In addition, radioactivity is mostly understood as a microscopic active entity, characterized as unstable, increasing, very strong and energetic, but light. It is also perceived as mobile and transient, although lasting; amorphous, yet dispersed and indivisible; immaterial, but detectable and measurable. At the same time, radioactivity is recognized as a special and rare entity, extraordinarily dangerous, destructive and harmful. Paradoxically, it is generally considered useful and, to a lesser extent, controllable. These contradictory appreciations are very likely due to the post-formal nature of the entity, which leads to build complex, relative, antithetical and multifaceted mental pictures of it. The data analysis has also revealed a progression in the understanding of radioactivity with formal education. On the one hand, the cluster analysis of Item 1 has shown a tendency to include radioactivity in the intermediate ontological categories `dispersed' and `immaterial' in compulsory secondary education. On the other hand, the higher levels place radioactivity with more certainty in the category `microscopic', indicating that the differences among ontological groups are better appreciated at increasing educational training. We have also observed discrepancies in the properties attributed to radioactivity as a function of the educational level. Here, trainee teachers are more aware of the relative and multidimensional character of radioactivity and show a more cohesive and unbiased view of the phenomenon. Accordingly, while secondary school students perceive it as strong, lasting, special, artificial, rare, wavy and increasing, the master group prefers to define it as both strong and weak, artificial and natural, wavy and corpuscular, brief and lasting, or increasing and decreasing depending on the context. Yet the most significant differences are found in the emotional dimension. While secondary education students have a biased view of radioactivity, showing uneasiness and restlessness towards it, the perspective of trainee teachers is more neutral. The former classify radioactivity as highly destructive, damaging and dangerous, whilst the later realize that it can be either destructive or creative, harmful or beneficial, and dangerous or safe depending on the situation. The observed change in the affective perspective might have its origin in the more extended scientific base of the master students. For instance, they generally know that radioactivity is far more common and frequent than believed by the broad audience or that its duration is an intrinsic property of each radionuclide. This improved understanding of the scientific facts related to radioactivity allows them to make fair-minded judgements.

\subsection{Item 3: on the perception of the knowledge about radioactivity}
\label{sec:perception}

\begin{figure}
\centerline{\includegraphics[width=0.55\textwidth]{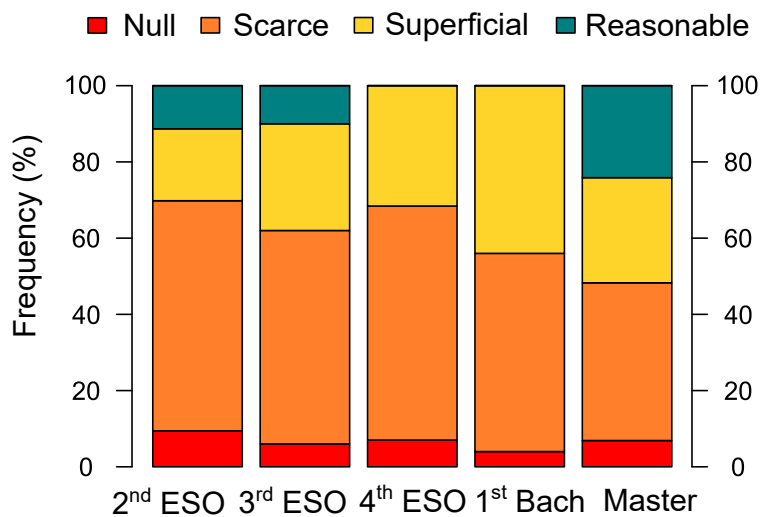}}
\caption{Response frequencies to Item 3 (`\emph{Your knowledge about radioactivity is ...}') as a function of the educational level of the inquired students. Percentages are normalized to the number of students of each level. The colour code of each answer is given in the legend.}
\label{fig5}
\end{figure}  

Figure \ref{fig5} reveals a clear agreement on what the inquired students think they know about radioactivity.  While the option `deep' is not selected at all, options `null' and `scarce' are the preferred ones by at least 50\% of students in all levels, with a gradually decreasing trend as the formal training rises. The progression of option `reasonable', instead, is quite erratic. In $2^{nd}$ ESO and $3^{rd}$ ESO, nearly 10\% of students believe their knowledge on radioactivity issues is `reasonable', while nobody in the higher secondary levels opts for this answer. This might be due to the fact that students with a lower level of formal education tend to overestimate their knowledge and, as a consequence, they are more likely to fall in an \emph{illusion of understanding} \cite{Tal10}. On the contrary, master students seem more cautious since, even if they have received advanced training on NS topics, only 20\% of them believe to have a `reasonable' knowledge. Finally, the low frequencies for the option `null' (below 10\% in all levels) support the idea that an important part of the radioactivity-related information comes from non-academic sources. In the following, we explore this possibility.  

\subsection{Item 4: On the sources of information}
\label{sec:sources}

\begin{figure*}
\centerline{\includegraphics[width=\textwidth]{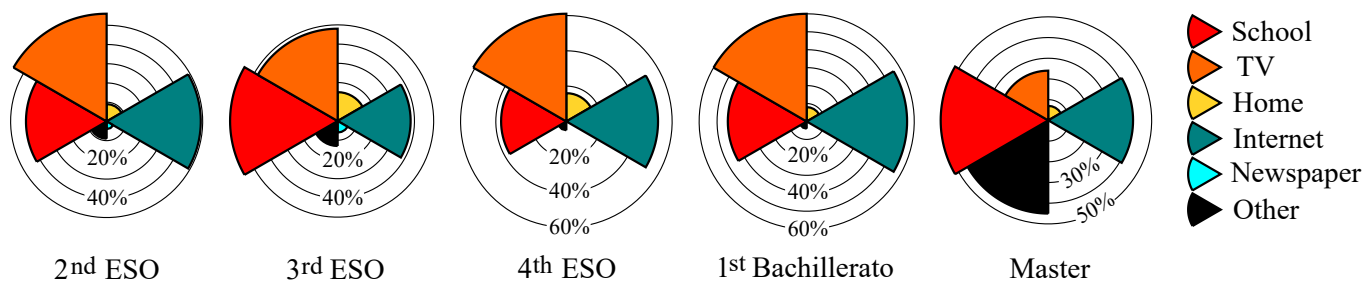}}
\caption{Selection frequencies for each answer given in Item 4 (\emph{Where have you acquired your knowledge on radioactivity?}). Results are shown separately for each academic level. The percentages are normalized to the total number of students in each educational stage.}
\label{fig6}
\end{figure*}

Item 4 poses the question `\emph{Where have you acquired your knowledge on radioactivity?}'. Figure \ref{fig6} shows the pie plots with the response distributions as a function of the educational level. At first sight, one can see the main information sources are `Internet', `TV' and `school'. In the case of `school', the larger frequencies are obtained for $3^{rd}$ ESO and the master group. Casually, the first educational level with radioactivity-related contents in the curriculum of Physics and Chemistry is $3^{rd}$ ESO, while the master students are the only group that has received the training on radioactivity foreseen in 2$^{nd}$ Bachillerato. Moreover, above 40\% of them declare having received advanced instruction on NS at undergraduate level, as indicated by the black pie in the right panel of Fig. \ref{fig6}, which corresponds to the answer `other'. Finally, the low frequencies of the items `home' and `newspapers' might be due to different factors, as a lack of awareness of the broad audience on the subject \textendash which can be related to the lack of a \emph{public understanding of Science} \cite{Mar92}\textendash~ or the increased use of digital communication technologies at the expense of the printed media.       

\subsection{Item 5: On the interests on NS applications}
\label{sec:interests}

\begin{figure}
\centerline{\includegraphics[width=0.6\textwidth]{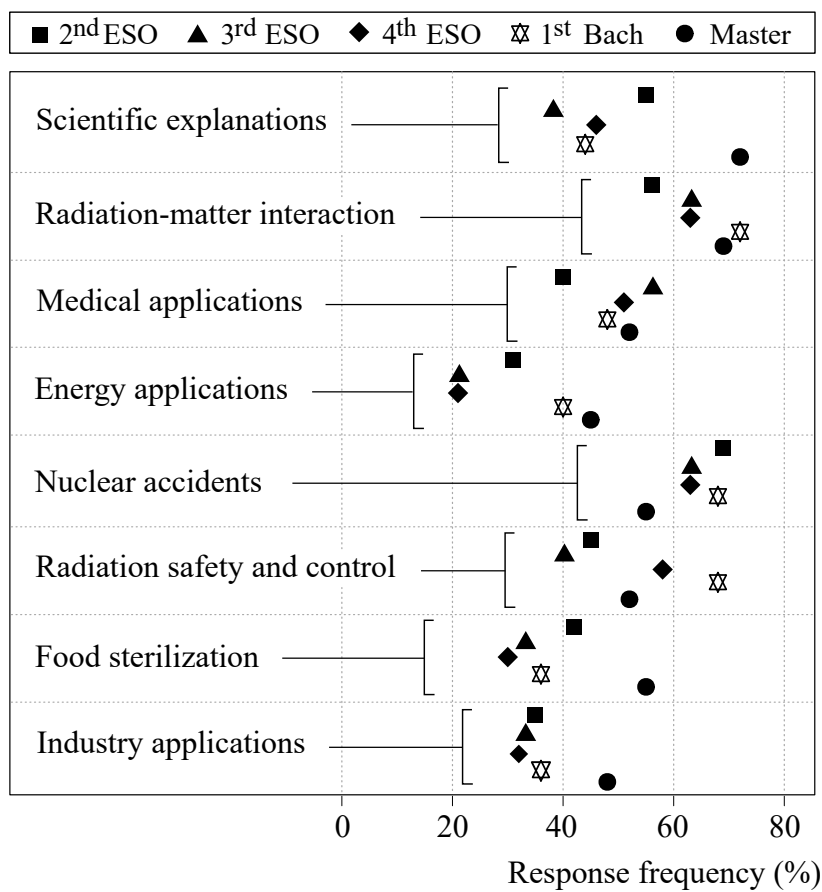}}
\caption{Selection frequencies for each answer given in Item 5 (\emph{Would you like to know more about any of the following aspects of radioactivity?}). Results are shown as a function of the academic level (see legend in the top panel). The percentages are normalized to the total number of students in each educational level.}
\label{fig7}
\end{figure}

Item 5 poses the multiple-choice question `\emph{Would you like to know more about any of the following aspects of radioactivity?}'. There are nine possible answers: `scientific explanations', `interaction of radiation with matter', `medical applications', `energy applications', `nuclear accidents', `radiation safety and control', `food conservation and sterilization', `industry applications', and `other'. Of these, individuals can select as many as they want. The results for the eight first answers are shown in Fig. \ref{fig7}, where the frequencies of response are given for each academic level. At first sight, one can see that the preferred topics are `nuclear accidents' and `radiation-matter interaction', with about 60\% of students showing their interest in the corresponding topics. This suggests that those aspects inspiring more curiosity in students, overall in secondary education, are the cause-effect relationships of radioactivity; in particular, its harmful effects on the human beings and the environment. These options are followed by `radiation safety and control', `scientific explanations' and `medical applications'. Finally, categories `food sterilization', `industry applications' and `energy applications' are at the bottom, with around 40\% of students wishing to know more about them. Even if there are no statistically significant differences among levels, a qualitative analysis of Fig. \ref{fig7} reveals that master students generally show a greater interest for the applications and scientific explanations of radioactivity. Such a progression appears to be logical since the higher the level of scientific literacy, the better the appreciation and enthusiasm towards science-related issues \cite{Sol97,Ace03}.    
 
\begin{figure}
\vspace{0.3cm}
\centering
\centerline{\includegraphics[width=0.55\textwidth]{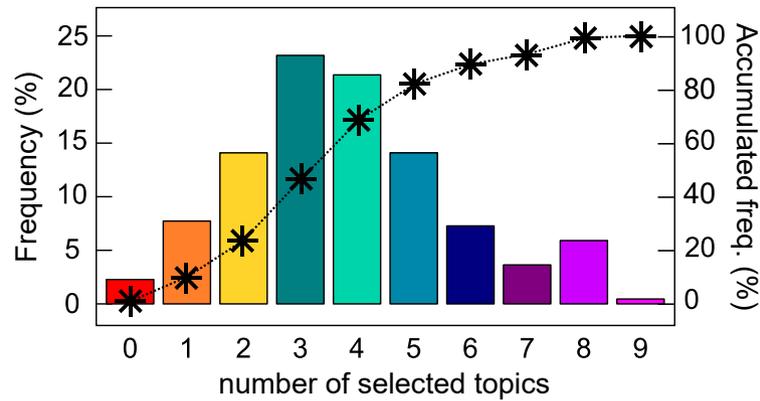}}
\caption{(Bars, left axis) Frequencies of the number of topics selected by students in Item 5. (Asterisks, right axis) Accumulated distributions of topics selected by students in Item 5. In both cases the frequencies are normalized to the total number of inquired students.}
\label{fig8}
\end{figure}

The number of selected topics per student has also been examined in order to asses the attention drawn. The results are shown in Fig. \ref{fig8}, where the normal (bars) and accumulated (asterisks) frequencies are displayed as a function of the number of aspects that each student has chosen. At first sight, one can see that only around 2\% declares no interest in any topic. Meanwhile, the figure reveals that well beyond 70\% selects more than 2 topics. We can conclude, then, that the willingness of students to learn STSE aspects based on radioactivity is excellent. A question to be answered now is whether their general knowledge on the subject is as good as the interest shown.

\subsection{Item 6: On the knowledge and misconceptions about radioactivity}
\label{sec:item6}

The aim of Item 6 is twofold. On the one hand, we pretend to assess the general knowledge on the NS contents foreseen in the Valencian curriculum of Physics and Chemistry; on the other, we aim at shedding light on the rationale behind the most widespread misconceptions about radioactivity exhibited by the inquired individuals.

\subsubsection{General knowledge of NS}
\label{sec:knowledge}

\begin{figure*}
\centerline{\includegraphics[width=\textwidth]{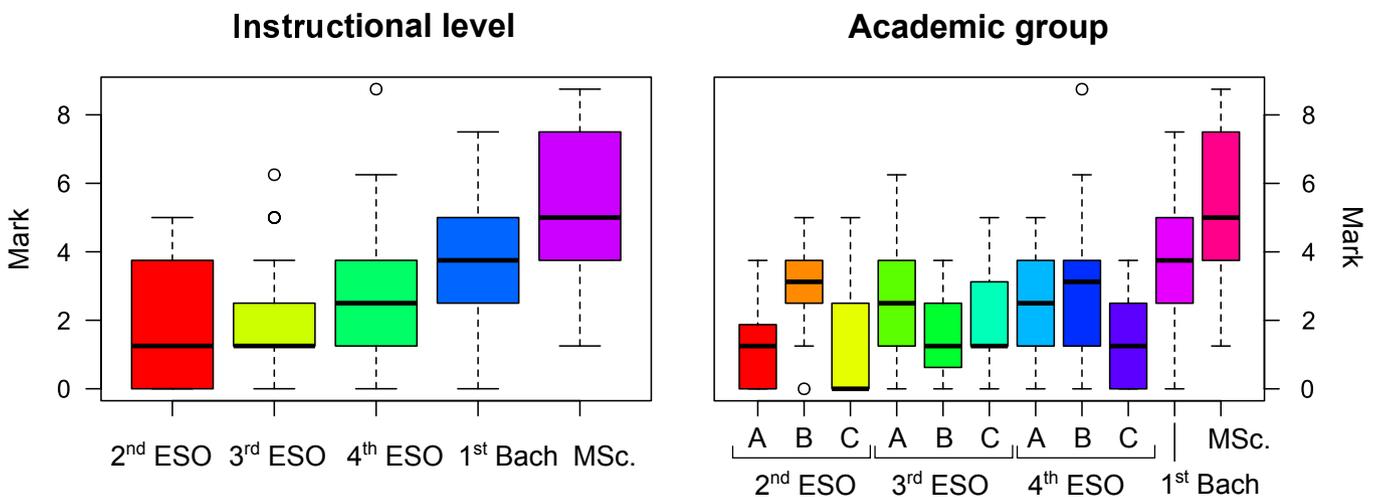}}
\caption{Marks obtained for Item 6 after applying the correction criteria described in Table \ref{table1}. The marks are shown as a function of the instructional level (left) and the academic group (right) of the surveyed individuals.}
\label{fig9}
\end{figure*}

The open-ended questions of Item 6 have been evaluated numerically by applying the rubric shown in Table \ref{table1}. The resulting marks are shown on a scale of 0 to 10 for each instructional level in the left panel of Fig. \ref{fig9}. As expected, the effect of formal education is highlighted by the increasing trend of the marks. Nonetheless, only the trainee teachers achieve on average the passing score. This suggests that most of the secondary school students have not achieved a meaningful learning of the NS concepts foreseen in their curricula. Given the interest shown by the surveyed students in the STSE aspects of NS (see previous section), their lack of knowledge could very likely be due to the scarce presence of Modern Physics in these curricula \cite{BOE,DOGV} rather than to the lack of interest of the students.

The asymmetry of some box plots in the left panel of Fig. \ref{fig9} suggests that the average total score does not only depend on the educational level, but also on the group of students. In the case of the master group, the asymmetry can be understood in terms of the varied academic background of the trainee teachers. However, the instructional background of the secondary school groups should be more homogeneous. For this reason, we have performed an ANOVA test \cite{Fis92} of the socio-demographic variables `level' and `group' associated to the secondary school sample. The results reveal statistically significant differences for the two variables, with $p<0.01$ in both cases. These can be appreciated in detail in the right part of Fig. \ref{fig9}, where the total scores are shown for each of the 11  surveyed groups. It is to note that different groups in the same educational level have been called `A', `B', and so on. 

The group with the higher average mark in compulsory secondary education is $2^{nd}$ ESO B. Casually, these students watched and commented a documentary about the Chernobyl nuclear accident in the Physics and Chemistry class shortly before filling in the questionnaire. This leads one to hypothesize that the low scores are more likely due to the dearth of educational resources employed to learn radioactivity rather than the complexity of the NS concepts, as adduced by some teachers \cite{Tsa13}. In conclusion, more time and better educational strategies would help implementing more efficiently the meaningful learning of the radioactivity-related concepts.  

\subsubsection{Misconceptions and conceptual mistakes}
\label{sec:misconceptions}

Questions Q6.1 (`\emph{How many types of nuclear radiation do you know?}') and Q6.2 (`\emph{Which one do you think is the most dangerous for human beings?}') try to approach the confusion generated by the terms `radioactivity', `radiation' and `radioactive material'. In absolute agreement with the scientific literature \cite{Kac87,Mil90,Mar92,Pra05,Neu12,Plo16}, this ambiguity is also appreciated in the individuals surveyed here. Many of them think that other types of radiation, such as microwave radiation, solar rays, ultraviolet rays, X rays, or even the sound have a nuclear origin. Consequently, they attribute  many applications based on these types of radiation to radioactivity. Among others, they cite the computer, the mobile, the tablet, the TV, the oven, the microwave, radiographies, ecographies, etc. They also identify some protection measures for non-ionizing radiation as protection measures for nuclear radiation. In some cases, these are recommendations  without scientific base that have been spread by the media. In summary, they tend to connect the ideas related to waves with radioactivity. As a result, they mix up the cause-effect relationships of radioactivity and believe that some nuclei are radioactive because they emit waves.

Another group of students alludes to radioactive substances such as uranium, plutonium or nuclear residues when they enumerate types of nuclear radiation, i.e., they confuse `radioactive material' with `radiation'. As well, toxic substances such as mercury or applications of radioactivity such as nuclear power plants are cited in their answers.      

Making the attempt to explain a microscopic phenomenon without the adequate academic training is a difficult task. The goal of question Q6.3 (`\emph{Why some nuclei are radioactive?}') is exploring the mechanisms used by students to describe a phenomenon as complex and abstract as radioactivity from their current mental picture of the world. Generally, they use well-known macroscopic concepts to search for an explanation. But they also use more familiar concepts already learned at school that are equally abstract and microscopic, such as the electricity or the fission and fusion reactions. In their answers, we find again the confusion generated by the cause-effect relationships of radioactivity, the search for analogies with physical, chemical or biological phenomena and properties, and the belief that radioactivity can only be produced in nuclear reactors. Even one individual makes reference to what seems to be a pseudo-scientific explanation of radioactivity (`it has like an aura'). These statements already exhibit some misconceptions, for instance that radioactivity is dangerous and harmful, it needs a propagation medium, it can only be produced in fission and fusion reactions, it has the same properties than electricity or it acts like a virus or bacteria \cite{Eij90}. Only a few students give academically rigorous answers such as `they are unstable' or `they have an irregular combination of neutrons and protons'. Of these, only one individual alludes to the synthesis of nuclei to provide an explanation (`it depends on the circumstances under it was created').  

Question Q6.4 (`\emph{How can we protect ourselves from a radioactive substance?}') does not only address the conceptual complexity related to the microscopic world, but also the vagueness of the terms `irradiation' and `contamination'. As a result, a considerable number of students believe that they can protect themselves from radiation damages avoiding contact with the radioactive substance. In their reasoning, they do not consider the exposure to nuclear radiation nor the dose received. This is, they speak about radioactivity as if it were a toxic product or pathogen. It is to note that even if most students allude to radiation suits, their twofold purpose \textendash to avoid contamination from radioactive material and protect from certain types of radiation\textendash ~is not clear to them. Moreover, most of them imagine a nuclear emergency crisis when confronted to this question, as if they only could be affected by radioactivity in these situations (`hide in a bunker', `abandon the city', etc.). The catastrophic view of radioactivity, related to hazard and destruction, is the predominant misconception observed here. Others emerging ideas are related to the artificial origin of radioactivity (`get away from nuclear power plants') or the association with a microbial illness (`put the contaminated objects in quarantine'). 

\begin{figure}[t]
\vspace{0.5cm}
\centerline{\includegraphics[width=\textwidth]{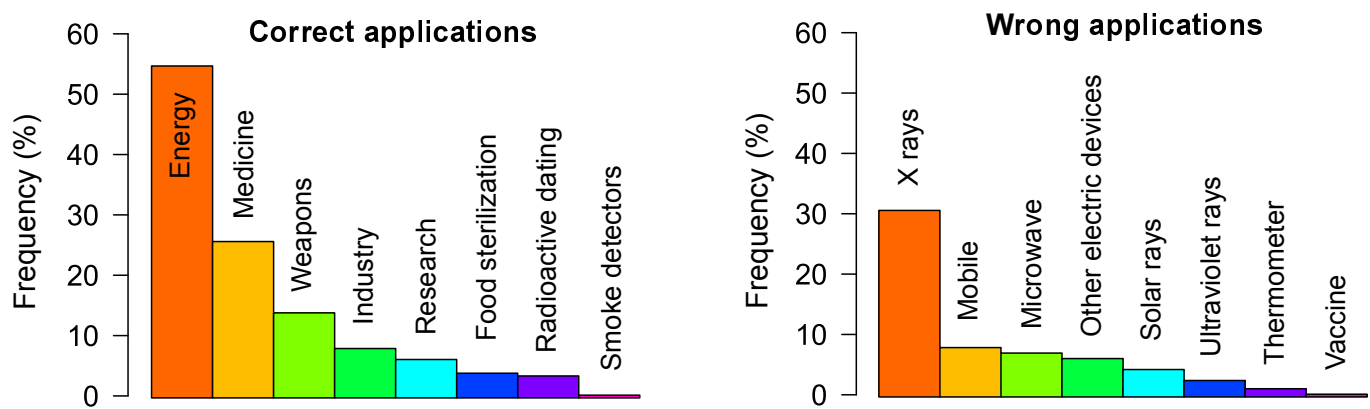}}
\caption{Applications of radioactivity cited in the answers to question Q6.5 of Item 6. Correct applications are shown in the left panel, while wrong applications in the right one. In both cases, frequencies are normalized to the total number of inquired students.}
\label{fig10}
\end{figure}

The aim of question Q6.5 (`\emph{How many applications of radioactivity do you know? List them.}') is exploring how well students have integrated the STSE relationships of NS in their mental schemes of radioactivity. Figure \ref{fig10} summarizes the results for this question. The left panel lists response frequencies for correct applications, while the right one indicates frequencies for wrong applications. Among the correct ones, we notice the production of electric energy, several applications of medicine, the development of nuclear weapons, industry applications, research, food conservation and sterilization and radioisotope dating. Only one master student cites smoke detectors of $^{241}$Am. In the bottom panel, six out of eight applications are based on electromagnetic radiations in frequency ranges lower than nuclear radiation. These are X rays, mobiles, microwaves, other electric devices, solar rays and ultraviolet rays. The other two applications are thermometers and vaccines. These provide further evidence of previously cited misconceptions and conceptual errors, such as the ambiguity caused by the terms `radiation', `radioactivity' and `radioactive material' or the connection established with toxic substances and viral or microbial diseases. 

In the figure, one can see that more than 50\% of students cite the production of nuclear energy. This high frequency is in line with the contents of NS foreseen in the Spanish and Valencian curricula of Physics and Chemistry (see Section \ref{sec:field}). Surprisingly, the second application most frequently cited are X rays, with about 30\% recurrence. This conceptual error can be motivated by different factors. For instance, the cognitive difficulty associated to the abstraction of the microscopic world, which contributes to mistake the atomic and nuclear domains. The closeness with some medical applications of X rays, such as the radiographies, might also be at the heart of this error. Or even some textbooks, which introduce  UV and X-ray ionizing radiations in the NS chapter without specifying their non-nuclear origin. As a result, it is not clear to students that only nuclear radiations can excite nuclei or induce nuclear reactions in their interaction with matter. Even though, the frequency of inquired subjects that do not list any erroneous application (54\%) almost doubles that of students that do not list any correct one (33\%).     

In order to build up the mental schemes that lead to a meaningful learning of radioactivity, a conceptual leap from the macroscopic to the microscopic world is required. A clear indication of such a fulfilment is given when the subject is capable of  identifying and differentiating chemical and nuclear processes. To this aim, question Q6.6 (`\emph{Do you think radioactivity affects living and inert matter equally?}') is posed. In general, students only consider the physiological effects of radioactivity in living matter. This, in some cases, leads to the erroneous conclusion that the structure of inert matter is not altered by radioactivity. Consistently,  a biological terminology is usually used in the answers, with references to genetic mutations and analogies with viral and microbial diseases. Among the response patterns, four groups are found. The first comprises subjects who believe that living beings become ill and die (`the duration of radiation is different. In an object it can last years, but it kills living beings over time'). The second comprises students that think radioactivity only transmutes the structure of living matter (`radioactivity changes the composition of human beings and causes their death. Meanwhile, inert matter does not change its morphology'). The third group believes that radioactivity affects more living matter \cite{Kla90,Als01} (`Radioactivity affects more living beings because it can react more easily with living matter than with lifeless objects'). Finally, the fourth group believes that only inert matter can be radioactive (`Radioactivity does not affect inert matter, but it can remain in it and infect living matter'). As a matter of fact, 26\% of students consider radioactivity can react equally with living and inert matter, yet only a few justify adequately their response (`both are equally formed by atoms').   

Question Q6.7 (`\emph{Can radioactivity turn objects radioactive?}') mainly addresses two misconceptions. The first is the confusion caused by the terms `irradiation' and `contamination' \cite{Mil96}. The second appears reiteratively in radioactivity teaching research and is related to the idea that objects become radioactive when they are irradiated \cite{Pra05,Nak06,Ses10}. Actually, such an idea has to be put into context, since matter can only become radioactive by irradiation with nuclear radiations or by contamination with radioactive material. As a matter of fact, only Ref. \cite{Plo16} clarifies this idea is a misconception for the whole electromagnetic spectrum excepting gamma radiation. In fact, it should be emphasized that the idea is true for most of the nuclear radiation (including $\alpha$ rays, $\beta$ rays, neutrons, etc.) because it is energetic enough to excite other nuclei or induce nuclear reactions that result in the production of unstable nuclei. Five response patterns are found here. The first  only makes reference to irradiation (`The properties of an object or organism can be altered in order to fulfil the conditions to be called radioactive'). The second cites contamination (`atoms of radioactive objects must be transferred to other objects'). The third does not consider the existence of nuclear reactions (`No, each element has its own features and properties'). The fourth group conceives the idea of radiation as `something' that can be conserved and accumulated in matter (`radioactive particles enter and remain in matter and that is how it turns radioactive'). Finally, the latter group associates radioactivity with illness (`if [radioactivity] is used in the human body, it can cause cancer or diseases').

In their answers, we find again difficulties to implement a microscopic nuclear model that explains nuclear reactions, most likely because radioactivity is an imperceptible, intangible, and abstract phenomenon that individuals cannot directly experiment. This, in turn, boosts the development of further misconceptions, such as that elements cannot transmute. The idea that radioactivity can be accumulated in matter, already cited in Ref. \cite{Eij90}, also appears here. Only one individual argues adequately `all nuclei, just because they are nuclei, can be radioactive'.     

\section{General remarks}
\label{sec:remarks}

The present study provides an excellent ontological frame to understand the nature and properties of radioactivity and its related ideas using a questionnaire as  diagnostic tool. It is possible then to inspect how the ideas about radioactivity develop and evolve along the different educational stages and determine when, how and why the alternative ideas appear in both the cognitive and affective domains. In general, the misconceptions and conceptual mistakes found are very similar to those already cited in the  literature. The most frequently observed here is the ambiguous use of the terms `radioactivity', `radiation' and `radioactive material'. A quick review of books, reviews and journals suggests that its most probable origin is non-academic. Furthermore, this misconception boosts the appearance of others, such as that the ozone layer can protect us from radioactivity or that the electrical devices are radioactive. But the most common one is the confusion caused by the terms `irradiation' and `contamination', which are related to `radiation' and `radioactive material' as indicated in Ref. \cite{Mil96}. Regarding the association with contamination \cite{Boy94,Neu12}, radioactivity is frequently related to environmental problems of differing origin such as the green-house effect, the hole in ozone layer or the global warming. As a curiosity, some students cite `toxic emissions from industrial plants', `cow farts' and `plastics and cans' as applications of radioactivity. 


The conceptual leap associated to the abstraction of the microscopic world also stimulates a number of misconceptions. One  widely spread is the confusion between the basic unit of nuclear matter, i.e. the nucleus, and the basic units of Chemistry and Biology, i.e. the atom and the cell. The lack of differentiation of the nuclear and atomic domains \cite{Pos89,Nak06} might have an historical origin since the discovery of radioactivity preceded that of the atomic nucleus and, as a consequence, scientists believed at the beginning of the XX century that radioactivity had an atomic origin. Nonetheless, some authors attribute this mistake to the confusion caused by the term `radiation' \cite{Kac87}, since both the atom and the nucleus are sources of electromagnetic radiation. In the present study, a significant percentage of students cite X rays as a type of nuclear radiation, and radiographies as an application of radioactivity (see Fig. \ref{fig10}). In line, some of them justify the nature of radioactivity through chemical processes ([nuclei are radioactive due to] `the number of electrons', `the atomic structure', `the chemical composition' or `the stability of electrons'). As a consequence, there are difficulties to distinguish the concepts of atom, nucleus, isotope and radioisotope \cite{Nak06}. As well, some find hard to believe that all chemical elements can become radioactive. Although in a lesser extent, the association with biological processes also emerges in some explanations (`radioactivity passes on from generation to generation through genes', `dead cells do not absorb radiation').     

The use of analogies with more familiar entities has also been extensively observed here. This strategy is commonly and unconsciously used by individuals to build up a cognitive bridge towards more complex and abstract concepts. In the case of radioactivity, the most noticed one is the analogy of the microbial disease \cite{Eij90}, strongly related to the idea that radioactivity affects living and inert matter in a different manner \cite{Kla90}. Indeed, the use of a medical vocabulary to make reference to radioactivity is quite usual in the responses to the open-ended questions. Terms such as `infection', `transmission', `illness', `cancer', `tumour', `toxicity', `vaccines', `virus', etc. frequently appear. Consistently, it is believed that once radiation enters into one body, it `remains' there until the body dies, making the comparison with a virus or bacteria. Others believe radiation is a tangible entity that can be `attracted' by atomic nuclei and `remain' in there, as if they were speaking of a toxic agent.     

Apart from the chemical and biological processes mentioned before, students also establish analogies with physical phenomena already learned at school, such as the sound or the electricity. They normally use macroscopic properties, such as the pressure, the mass or the volume, to explain radioactivity. They also use known microscopic concepts such as the fission and fusion reactions, although in a lesser extent. It is hard to find references with similar misconceptions in the scientific literature. Perhaps the closest one is the belief that radioactivity is produced in nuclear reactors \cite{Boy94}, as fission reactions are implicitly used there to build a coherent explanation of the phenomenon.  Similarly, some students relate the origin of radioactivity to nuclear fuel.

The danger associated to a large time exposure to nuclear radiations or the time needed for radioactivity to disappear are also often brought up by students. These ideas are intimately related to concepts still unknown in compulsory secondary education, such as the half-life, the radioactive dose or the radioactive decay rate. Some students believe that a prolonged or intense exposure to radiation turns bodies radioactive, no matter what the type of radiation is. Others insist in the persistence of radioactivity through the microbial disease analogy. In this case, it is perceived as a material and tangible entity that can be accumulated and transferred to other bodies. As such, radioactivity is believed to remain in matter unless it comes out or it is somehow extracted. For some students, this misconception has evolved into a technological version in which bodies can `accumulate' radioactivity, as if they were batteries than can be charged and discharged. This analogy does not seem to have been detected before in the scientific literature, most likely because this is one of the first systematic enquiries on radioactivity to digital natives. The idea that atoms do not change their nature, or that elements are immutable, also appears linked to the notion of durability. 

In the present study, the only nuclear reactions mentioned by the secondary school sample are the fusion and fission reactions, precisely the only two foreseen in the Valencian curriculum of Physics and Chemistry. This suggests an unawareness of the existence of other types of nuclear reactions in nature and, at the same time, explains why most of the secondary students attribute an artificial origin to radioactivity.

Finally, it should be noted that the attitudes and emotions inspired by radioactivity are approached in some items of the questionnaire, inspecting which applications do students know and which is their perception about their utility, advantages and disadvantages. As a result, secondary school students identify radioactivity with danger, damage, illness, contamination and destruction, showing a biased perspective. Despite recognizing some beneficial applications such as medicine, they generally exhibit an instinctive fear to radioactivity, most likely due to the catastrophic view given by Internet and the mass media, which bring death into focus.

\section{Conclusion}
\label{sec:conclusion}

 We have carried out the first systematic study about misconceptions, knowledges and attitudes of students in educational centres of the Valencian Community (Spain). The work, carried out during spring 2018, has verified most of the misconceptions and conceptual errors reported in the scientific literature. Given the new intervention context, some up-to-date nuances have been observed and discussed. There has only been one previous misconception unidentified in the present work, `radiation can be used to detect feelings'  \cite{Neu12}. On the other hand, the knowledge that students have on the basic aspects of radioactivity is clearly insufficient  to properly establish the STSE relationships of NS. This aspect seems fundamental to develop the emotional facet of radioactivity, which shows a clear evolution from a radical catastrophic view in the compulsory levels to a more moderated, critical-thinking based perspective in the master group. Regarding the attitudinal dimension, we have verified a strong interest towards the applications of NS, with an overwhelming 98\% of students wishing to learn new aspects. As shown in Fig. \ref{fig7}, nuclear accidents and effects of ionizing radiations in living a inert matter raise the highest interests in secondary education, whilst more technical issues, such as the scientific explanations of radioactivity or its applications, are better appreciated at increasing instructional level. Importantly, the results highlight the need to develop new teaching strategies to approach the deficiencies found here. Such procedures must lead to a meaningful learning of the NS concepts in order to promote the critical-thinking skills necessary to discuss and make decisions about the most controversial STSE aspects of nuclear science.





\subsection*{Conflict of interests}

The authors declare no potential conflict of interests.

\section*{Supporting information}

The following supporting information is available as part of the online article:

\noindent
\textbf{Questionnaire}

\bibliography{bibliografia}

\end{document}